\documentclass[reprint,amsmath,amssymb,aps,pre,longbibliography]{revtex4-2}
\usepackage{amsmath}
\usepackage{float}
\usepackage{graphicx}
\usepackage{subfigure}
\usepackage{dcolumn}

\usepackage{amssymb}
\usepackage{amsmath}
\usepackage{bm}
\usepackage{graphicx}					

\usepackage{version}				
\usepackage{cases}						

\usepackage{mathrsfs}

\usepackage{graphicx}
\usepackage{xcolor}
\usepackage{psfrag}
\usepackage{enumerate}
\usepackage{amssymb}
\usepackage{amsmath}
\usepackage{wasysym}
\usepackage{epstopdf}	
\usepackage{ulem}

\newcommand{\ew}[1]{\textcolor{black}{#1}}
\newcommand{\xh}[1]{\textcolor{black}{#1}}

\begin{document}

\title{\ew{Sharp depletion of radial distribution function of particles due to collision and coagulation inside turbulent flow: a systematic study.}}

\author{Xiaohui Meng}

\author{Ewe-Wei Saw}
\email{ewsaw3@gmail.com}
\altaffiliation[Also at ]{Ministry of Education Key Laboratory of Tropical Atmosphere-Ocean System, Zhuhai, China}
\affiliation{School of Atmospheric Sciences and Guangdong Province Key Laboratory for Climate Change and Natural Disaster Studies, Sun Yat-Sen University, Zhuhai, China}


\begin{abstract}

\ew{The clustering (preferential concentration) and collision of particles in turbulent flow is a significant process in nature and technical applications. We perform direct numerical simulation (DNS) to study the clustering of small, heavy, monodisperse particles subject to collision-coagulation in turbulent flow (i.e., colliding particles always coagulate (coalesce) into large ones). We find that collision-coagulation causes the radial distribution function (RDF) of the particles to decrease strongly at particle separation distances $r$ close to the particle diameter $d$. {However, we observe that the RDF do not decrease indefinitely but approach a finite value in the limit of $r\to d$.} We study how the characteristics of this ``depletion zone" relate to the particle Stokes number (St), particle diameter, and the Reynolds number of the turbulent flow. 
A collision-induced modulation factor $\gamma_{c}$ is defined to represent the degree of RDF depletion due to collisions-coagulation.}
\ew{In the region where $\gamma_c(r)$ is a quasi-power-law, we find that the corresponding power-law exponent $\tilde{c}_1$ only depends weakly on $St$. We also find that the overall trend of $\tilde{c}_1$ with respect to $St$ is similar to that of the classical power-law exponent $c_{1}$ appearing in the RDF of non-colliding particles, i.e., the exponent increase at small $St$, peak around $St \approx 0.7$, and decrease thereafter. The same qualitative trend is also observed for the limiting values of $\gamma_c$ at $r\to d$.} 
\ew{A complementary investigation on the Stokes number trend of the full RDF in the depletion zone is conducted. The slope of RDF appear constant for $St\ll1$ but is changing when $St$ is getting large. \xh{The position where the RDF starts to decrease is found to be $St$-dependent.}}
\ew{We found that he depletion zone is insensitive to the flow Reynolds number and $\gamma_c$ of different $Re_{\lambda}$ overlap. With changing particle diameter $d$, the reduction of RDF occurs at scales that shift accordingly and always starts at around $2.4d-3d$. We also find that the shape of $\gamma_c(r)$ is independent of changes in $d$.}
\end{abstract}

\maketitle

\section{\label{sec:level1}Introduction}

The fluctuation of particle concentration has a profound effect on inter-particle collision. This effect has a pivotal role in both natural sciences and industrial engineering. \ew{For example, the collision-coagulation process for small droplets (particles) determines their spatial and temporal size distribution}. These microscale properties have significant influences on the formation of precipitation \cite{shaw2003particle}. Small-scale turbulence in clouds makes an important contribution to the collision and coagulation of droplets \cite{grabowski2013growth} \cite{shaw2003particle}. \ew{Another} example is the formation of planetesimal. The collision of dust in protoplanetary disks sets the stage for planet formation. Research shows that the turbulent motion will concentrate dust in the dissipation \ew{scale, increasing the collision rate of dust particles \cite{cuzzi2001size} \cite{johansen2007rapid} \cite{pan2011turbulent}.} 

The preferential concentration of inertia particle \ew{has become a prevalent research topic since the end of the 20th century. Squires and Eaton \cite{squires1991preferential} found that the inertial particles concentrate preferentially in regions of low vorticity and high strain rate. The degree of particle clustering can be characterized via the radial distribution function (RDF), which is defined as} the ratio of the probability of finding a particle pair at a distance \ew{$r$} normalized by the probability \ew{of the same event} for random particle distribution. Reade \& Collins \cite{reade2000effect} found that the clustering of \ew{small} particles occurs on a scale that is much smaller than the Kolmogorov length scale and the RDF of particles follows a negative power law with the \ew{inter-particle} separation distance $r$. \ew{Chun et al.} \cite{chun2005clustering} developed a drift-diffusion theory to predict the RDF in turbulent flows for particles \ew{in the limit of small particle Stokes number. Their} results indicate \ew{that RDF} of particle is proportional to $c_{0}(r/\eta)^{c_{1}}$, where $\eta$ is the Kolmogorov length scale. They also find that the exponent $c_{1}$ is proportional to $St^{2}$. The Stokes number (St) is an important measurement for particle inertia, \ew{and is defined as} the ratio of particle relaxation time $\tau_{p}$ and the Kolmogorov time scale $\tau_{\eta}$. The dissipation-scale clustering of particles becomes stronger as the Stokes number increase and the RDF reaches a peak near the Stokes number of order unit \cite{ireland2016effect}. \ew{This relationship between the RDF and the Stokes number is corroborated by both numerical and experimental studies \cite{saw2008inertial} \cite{salazar2008experimental}.}

The preferential concentration of particle \ew{is expected to enhance  particle collision but the calculation of collision kernel is still an open question.} In the work of \ew{Sundaram and Collins} \cite{sundaram1997collision}, the RDF \ew{was} first introduced into the formula of collision kernel \cite{wang2000statistical}: $K=4\pi d^{2}g(d)\langle W(d)\rangle$, where $d$ is particle diameter, $g(d)$ is the value of RDF at $r=d$ and $\langle W(d)\rangle$ is the mean radial relative velocity of particle at $r=d$. \ew{The works on the RDF of particles mentioned thus far had used the ghost particle assumption and ignored the effect of collision and coagulation among particles. Saw and Meng \cite{saw2022intricate} found that the RDF will drop profoundly} at $r$ close to $d$ \ew{in the presence of collision-coagulation.} \ew{This finding is interesting because it highlights the importance of accounting for actual particle collision and it also questions the formula of collision kernel mentioned above.} 

In this paper, we use direct numerical simulation (DNS) to study the RDF of the \ew{inertial, colliding,} particles. \ew{DNS, which solves the Navier-Stokes equation fully resolving the spatial and temporal scales of the problem without using any turbulence modeling,} is an efficient numerical tool to study the particle-laden turbulent flow. The RDF of particle considering the effect of the collision-coagulation is investigated. The influence of the particle and turbulent parameters on the decreasing of RDF is also studied in this paper. The paper is organized as follows: section 2 provides a summary of simulation methods and the relevant turbulent and particle parameters. The statistical results and discussion are in section 3 and section 4. Finally, the results of the influence of the turbulent and particle parameters on the RDF are summarized in section 5.

\section{\label{sec:level1}Simulation method}

We performed direct numerical simulation (DNS) of the particle-laden turbulent flow. The incompressible Navier-Stokes equations are shown below.

\begin{equation}
	\frac{\partial\vec{u}}{\partial t}+\vec{u}\cdot\nabla\vec{u}=-\frac{1}{\rho}\nabla p+\nu\nabla^{2}\vec{u}+\vec{f}(\vec{x},t)
\end{equation}

\begin{equation}
	\nabla\cdot\vec{u}=0
\end{equation}

The N-S equations are numerically solved on $N^3$ grids cube using a pseudo-spectral method on the periodic domain, the length of which is $2\pi$. The turbulent velocity $\vec{u}$ is transformed from physical space to wavenumber space. The aliasing error rising from the convection part of N-S equation is removed by the 2/3-method  \cite{rogallo1981numerical}. $\vec{f}(\vec{x},t)$ in the N-S equation is an external forcing conducted at low-wavenumber in order to maintain statistically stationary \cite{eswaran1988examination}. \ew{In order to study the influence of the (Taylor scaled) Reynolds number on RDF, simulations with different $Re_{\lambda}$ are conducted}: $Re_{\lambda}=84$, $124$ and $189$, the detailed turbulent parameters are shown in Table \ref{tab1}. For different Reynolds number, the grid size is $N=256^{3}$ (for $Re_{\lambda}=84$ and $124$) and $512^{3}$ (for $Re_{\lambda}=189$). \ew{The small scale resolution may be characterized by $k_{max}\eta=1.59$, 1.21 and 1.38 respectively, where $k_{max}=N\sqrt{2}/3$ is the maximum resolved wavenumber magnitude.} The 2-order Runge-Kutta method is used to conduct time advancement in N-S equation. The Courant number $C=0.0248$, $0.0401$ and $0.0865$.

\begin{table*}
	\caption{The DNS parameters and time-averaged statistics. $N$ is the \ew{simulation grid size}, $\nu$ is the \ew{kinematic} viscosity of turbulence, $\epsilon$ is the dissipation rate of turbulent flow, $u^{\prime}$ is \ew{the} root-mean-square velocity of turbulent flow, $\lambda$ is the Taylor length scale, $\eta$ and $\tau_{\eta}$ are the Kolmogorov length and time scale, \ew{$L$ and $T_{L}$ }are the integral length and time scale, $Re_{\lambda}$ is the Taylor scaled Reynolds number.}
	\begin{ruledtabular}
	\begin{tabular}{c|cccccccccc}
		& $N$ & $\nu$ & $\epsilon$ & $u^{\prime}$ & $\lambda$ & $\eta$ & $\tau_{\eta}$ & $L$ & $T_{L}$ & $Re_{\lambda}$ \\
		flow 1 & 256 & 0.001 & 0.0326 & 0.3519 & 0.2386 & 0.0132 & 0.1750 & 0.5073 & 1.4416 & 84 \\
		flow 2 & 256 & 0.001 &  0.1013 & 0.5684 & 0.2187 & 0.0100 & 0.0993 & 0.6151 & 1.0822 & 124 \\
		flow 2 & 512 & 0.001 &  0.9472 & 1.226 & 0.1544 & 0.0057 & 0.0325 & 0.7398 & 0.6034 & 189 \\
	\end{tabular}
	\end{ruledtabular}
	\label{tab1}
\end{table*}

The particles we take are small (the diameter of particle $d$ is smaller than the Kolmogorov length scale $\eta$) and heavy (the particle's density is larger than the flow's). The gravitational effect and inter-particle hydrodynamic interactions are not included in our DNS model because only basic questions are discussed in this paper. Under these circumstances, the particles are only subjected to viscous Stokes drag force, and the motion equation of particles is shown below \cite{maxey1983equation}:

\begin{equation}
	\frac{d\vec{v}}{dt}=\frac{\vec{u}-\vec{v}}{\tau_p}
\end{equation}

In which, $\vec{v}$ is \ew{the particle velocity}, $\vec{u}(\vec{x},t)$ is \ew{the fluid velocity} at \xh{particle position}. $\tau_{p}$ is the particle inertia response time, defined $\tau_{p}=\frac{1}{18}\frac{\rho_{p}}{\rho}\frac{d^{2}}{\nu}$ where $\rho_{p}$ and $\rho$ are density of particles and flow respectively, $d$ is \ew{the particle diameter} and $\nu$ is dynamic viscosity of turbulent flow. The linear interpolation method is used for interpolating flow's velocity in particle position and the 2-order Runge-Kutta method with ``exponential integrators" is used for time advancement \cite{ireland2013highly}.

Spherical and mono-dispersed particles are randomly introduced in the simulation. Particles collide when their volumes overlap and a new particle is formed conserving volume and momentum. New particles are injected continuously and randomly in the system so that the particle system is in a steady state after \ew{a transient period.} In order to study the influence of particle parameters on RDF, particles with different Stokes numbers from $0.01$ to $2.0$ are introduced in each simulation. 
The size of particle is another important parameter related to particle collision. Three different sizes of particles are introduced in each simulation: $d=\frac{1}{3}d_{*}$, $d=d_{*}$ and $d=3d_{*}$, $d$ is the diameter of particle and $d_{*}=9.4868\times10^{-4}$. The details of simulations are listed in Table \ref{tab2}. 
\ew{The statistics are calculated for mono-dispersed particles.}

\begin{table}[H]
	\centering
	\caption{Characteristics of the runs discussed here. $Re_{\lambda}$ is the Taylor microscale Reynolds number of \ew{the} flow. $d$ is \xh{the particle diameter}, $d_{*}=9.4868\times10^{4}$. $St$ is \xh{the particle Stokes number}.}
	\begin{ruledtabular}
	\begin{tabular}{c|ccc}
		Run & $Re_{\lambda}$ & $d$ & $St$ \\
		1  & 124 & $d_{*}$ & 0.01 \\
		2  & 124 & $d_{*}$ & 0.05 \\
		3  & 124 & $d_{*}$ & 0.1 \\
		4  & 124 & $d_{*}$ & 0.2 \\
		5  & 124 & $d_{*}$ & 0.5 \\
		6  & 124 & $d_{*}$ & 0.7 \\
		7  & 124 & $d_{*}$ & 1.0 \\
		8  & 124 & $d_{*}$ & 2.0 \\
		9  & 124 & $\frac{1}{3}d_{*}$ & 0.1 \\
		10 & 124 & $3d_{*}$ & 0.1 \\
		11 & 84 & $d_{*}$ & 0.1 \\
		12 & 189 & $d_{*}$ & 0.1 \\
	\end{tabular}
	\end{ruledtabular}
	\label{tab2}
\end{table}

\section{Results and Discussion}

\subsection{Stokes number dependence}

\begin{figure}[htb]
\subfigure[]{
\label{Fig.rdf_st_part1}
\includegraphics[width=0.45\textwidth]{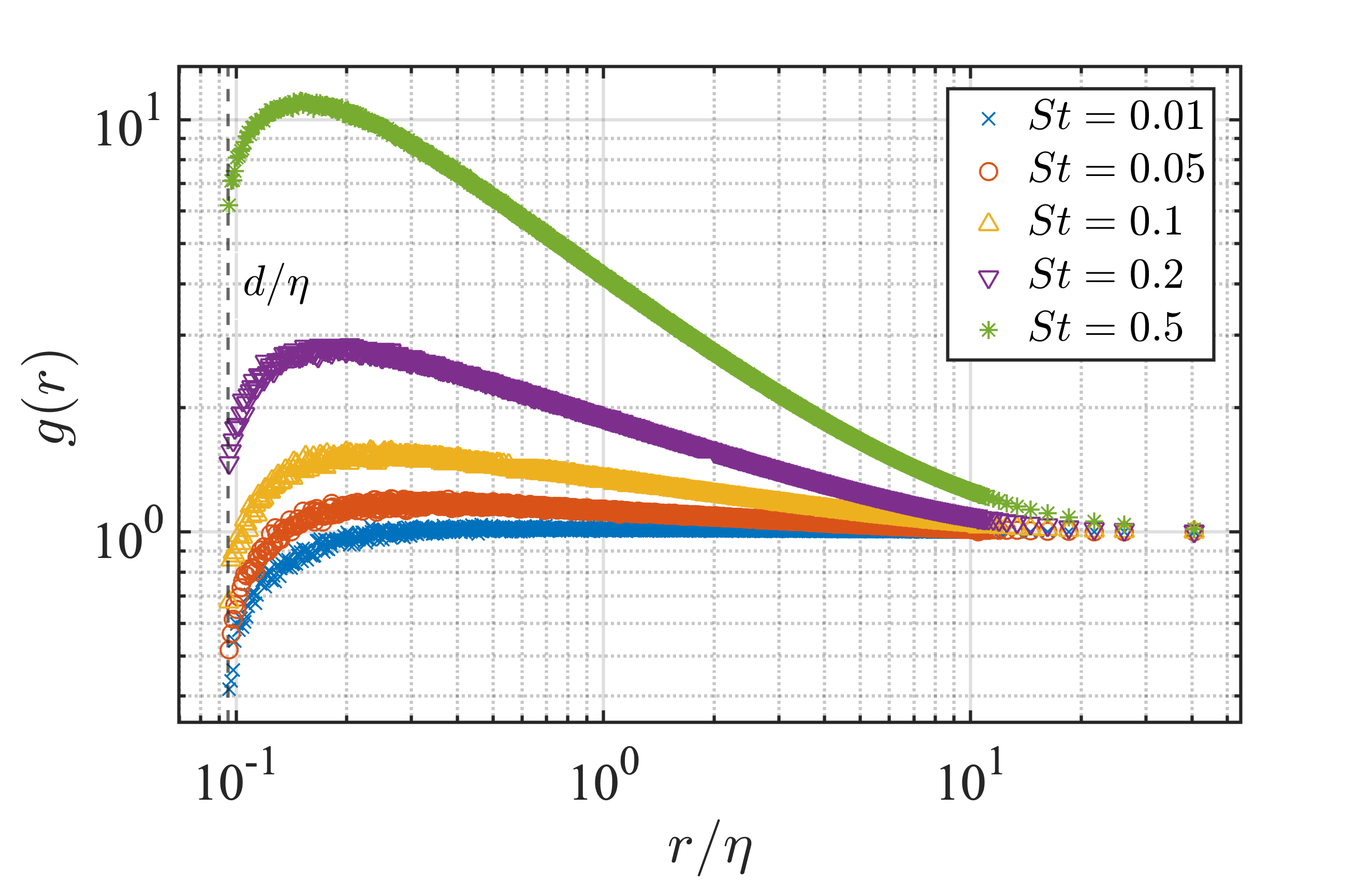}
}
\subfigure[]{
\label{Fig.rdf_st_part2}
\includegraphics[width=0.45\textwidth]{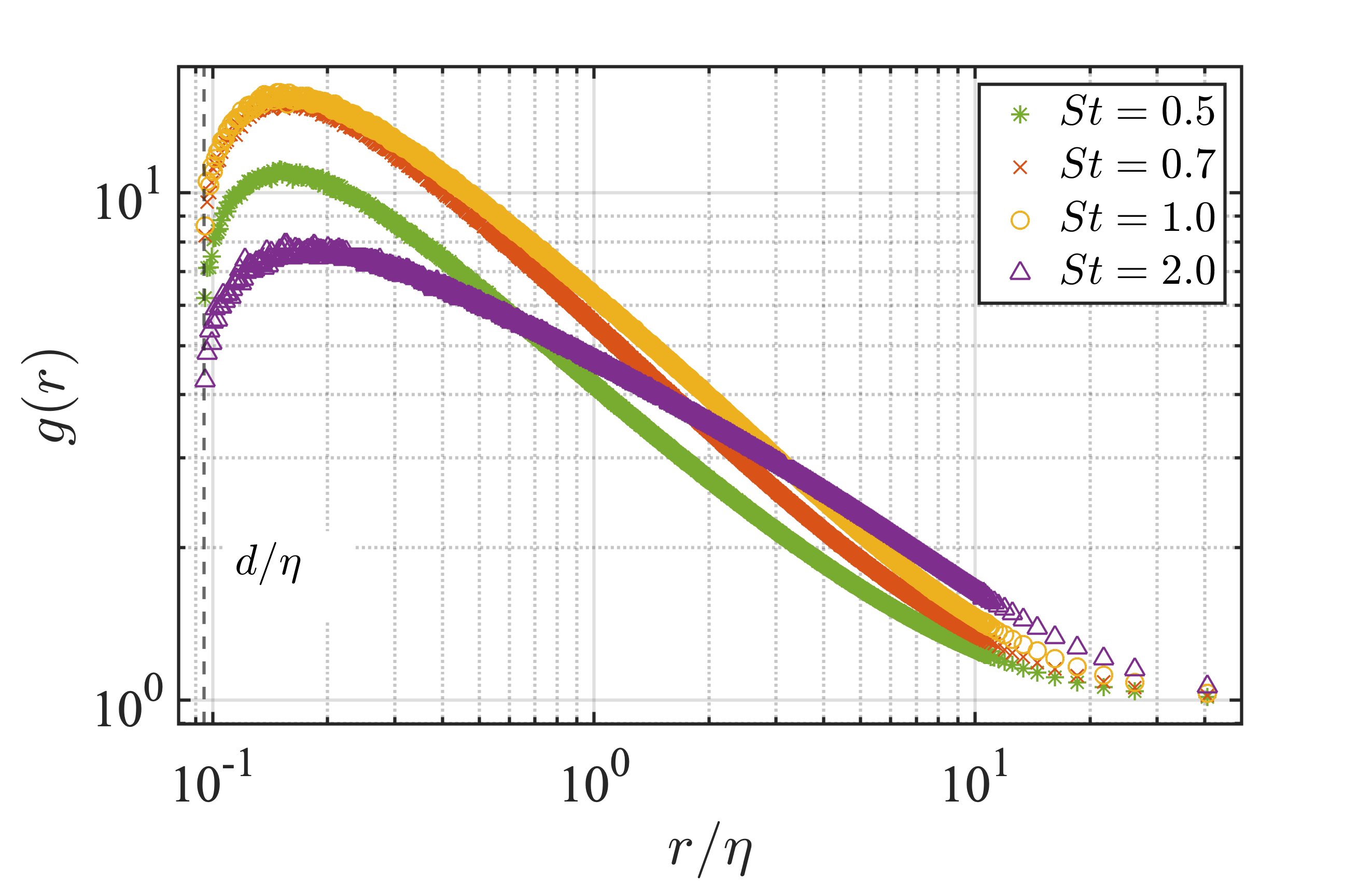}
}
\caption{{The RDFs versus $r/\eta$ for particles with different Stokes number. The diameter of particle is $d=9.49\times10^{-4}$ and the Taylor scaled Reynolds number is $Re_{\lambda}=124$. The RDF drops significantly when $r$ is close to $d$.}}
\label{Fig.rdf_st}
\end{figure}

\begin{figure}[htb]
\centering
\subfigure[]{
\label{Fig.rdf_st_d_part1}
\includegraphics[width=0.45\textwidth]{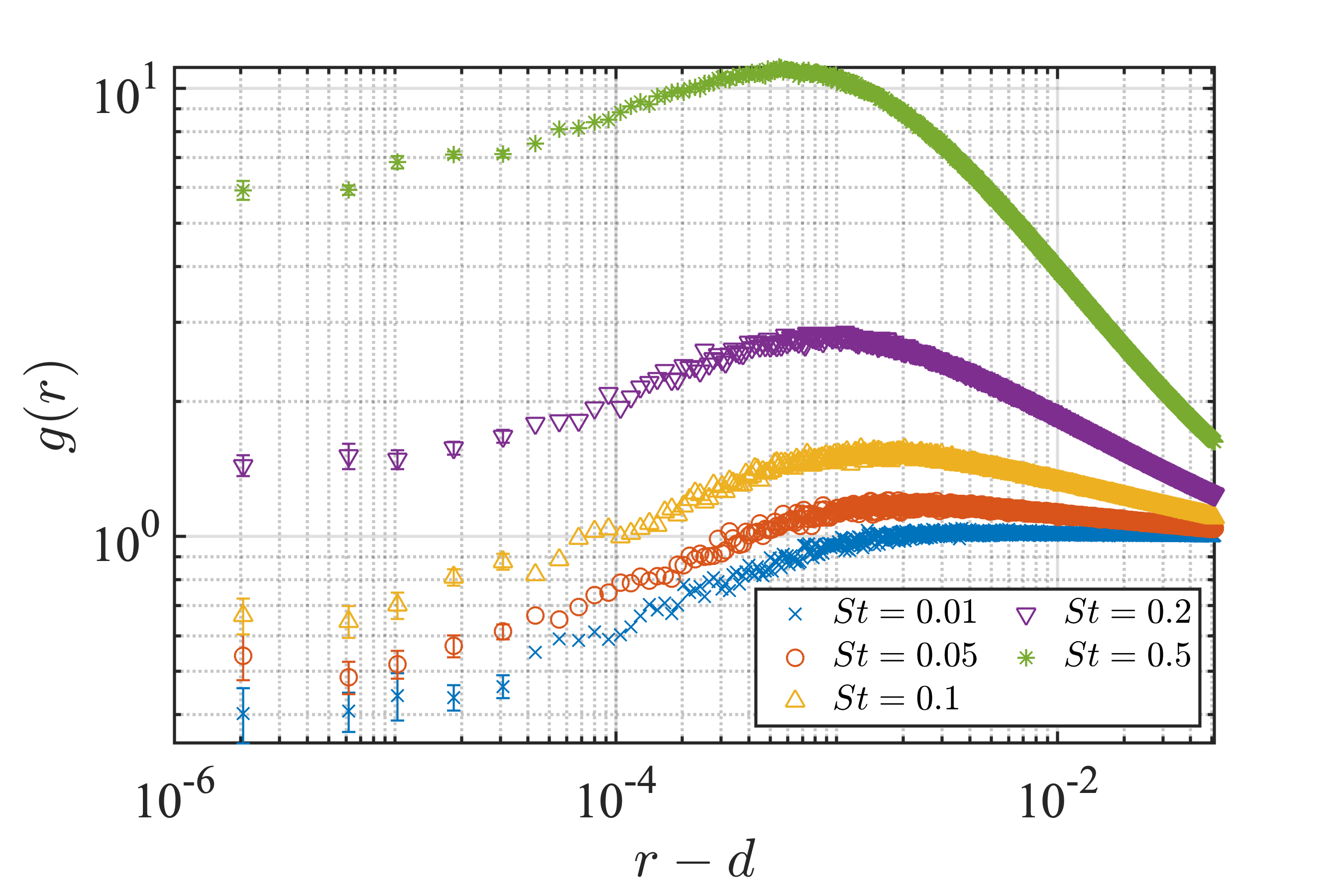}
}
\subfigure[]{
\label{Fig.rdf_st_d_part2}
\includegraphics[width=0.45\textwidth]{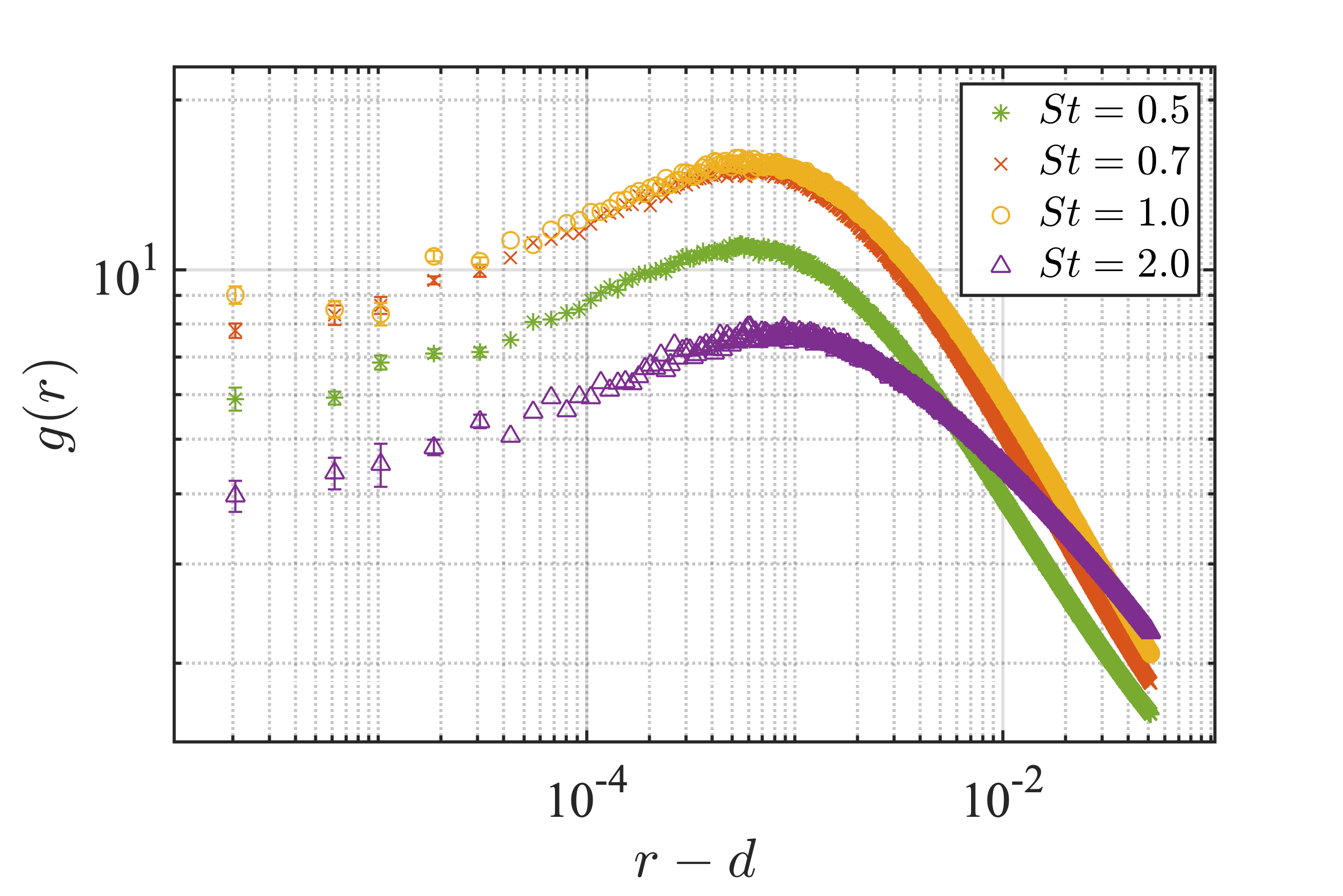}
}
\caption{The RDFs versus $r-d$ for particles with different Stokes number, $d=9.49\times10^{-4}$ is the particle's diameter. The Taylor-scaled Reynolds number is $Re_{\lambda}=124$ in this case. Error bars represent one standard error. The RDF follows a quasi-power-law with $r-d$ in the range $0.04d\lesssim r-d\lesssim0.3d$ and the slope of RDF decreases to zero gradually as $r$ continues to decrease.}
\label{Fig.rdf_st_d}
\end{figure}

\ew{The statistics from Run 1 to Run 8 in Table \ref{tab2} are used to study the influence of Stokes number on RDF. The RDFs for particles with different Stokes numbers are shown in Figure \ref{Fig.rdf_st}. What is striking in this figure is the strong decrease of RDF when the particle separation distance $r$ is close to the particle diameter $d$. Figure \ref{Fig.rdf_st} also show that there is an increase in the slope and the magnitude of RDF at the scales $r/\eta\sim1-10$ when $St$ is increased from 0.01 to 0.7. {We note that the slope of  the RDF for the $St=1.0$ case is smaller than that of the $St=0.7$ case, even though its values are everywhere larger.} Beyond $St=1.0$, the slope {and magnitude are both} decreasing from $St=1.0$ to $St=2.0$. At larger scales ($r/\eta \sim10$), the RDF curves flatten and converge to 1. {To elucidate the trend of RDF when the separation distance $r$ is close to the particle diameter $d$, the RDFs are plotted as the function of $r-d$ in Figure \ref{Fig.rdf_st_d}.} 
The most interesting aspect of this graph is that the relationship between RDF and $r-d$ exhibits a quasi-power-law trend in the range} \xh{$4\times10^{-5} \lesssim r-d \lesssim 3\times10^{-4}$}. \ew{As $r$ continues to decrease towards the particle diameter, the slope of RDF gradually becomes smaller and approaches zero, i.e., the RDF approaches a plateau.}

\begin{figure}[htb]
	\centering
	\includegraphics[width=0.45\textwidth]{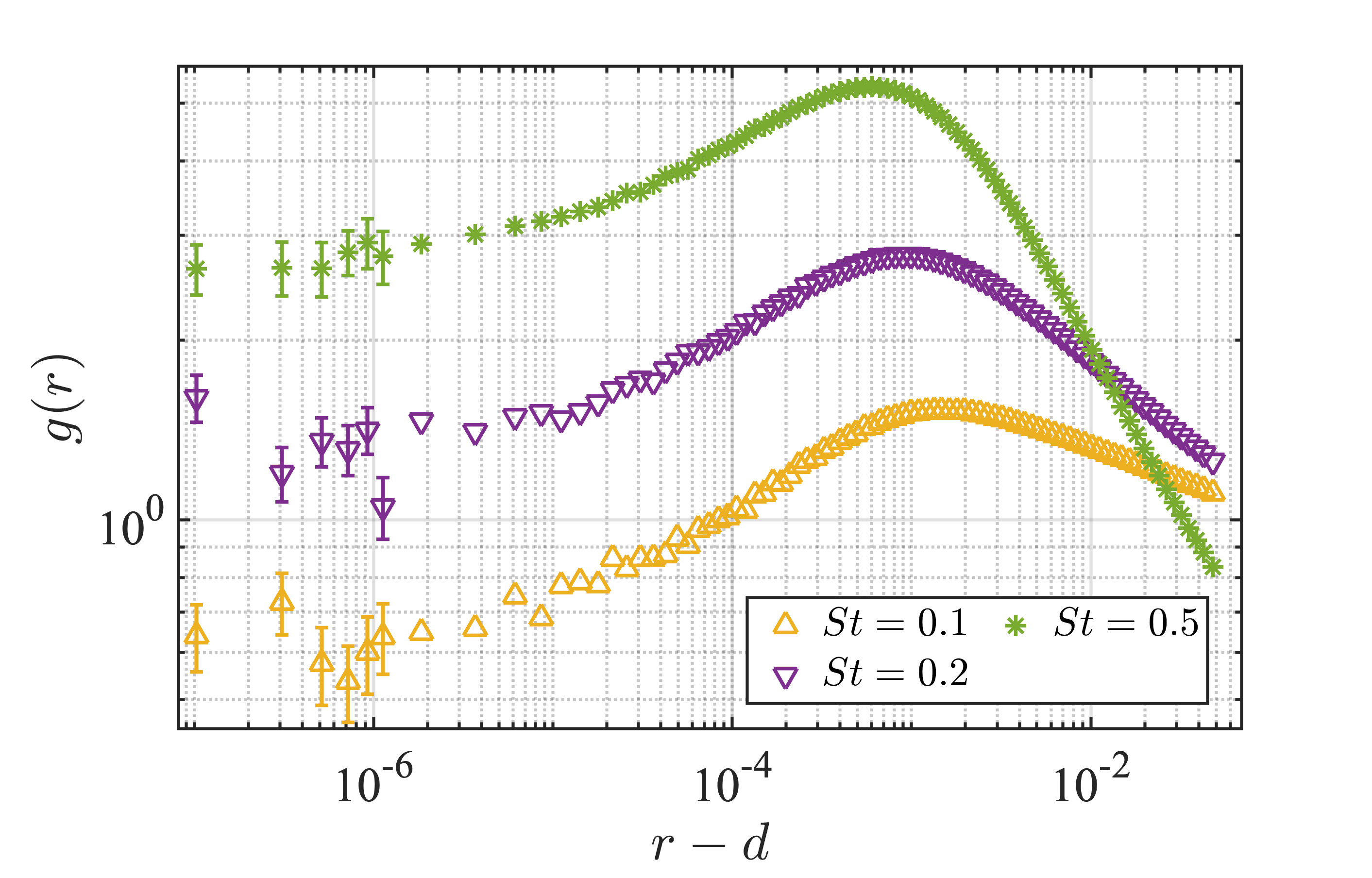}
	\caption{ {High resolution plots of RDFs versus $r-d$ for cases of $St=0.1$, $0.2$, and $0.5$, where the regime of smaller $r-d$ values is resolved more clearly. \xh{In order to compare them clearly, the RDF for $St=0.5$ is translated down vertically to half of its original height} Error bars represent statistical error of one standard deviation. Within the range of uncertainty, the RDF no longer decreases after $r-d<0.04d$ and levels off to a fixed value.}}
	\label{Fig.rdf_small_scale}
\end{figure}

\ew{Figure \ref{Fig.rdf_small_scale} illustrates more closely the trend of RDF as $r$ decrease towards $d$, using the cases of \xh{$St=0.1$, $St=0.2$, and $St=0.5$} as examples. This shows more clearly that after $r-d$ is less than $4\times10^{-6}$, RDF no longer decreases but levels off to a fixed value. 
The limiting value of the RDF at particle contact ($r=d$) is proportional to the particle collision rate \cite{sundaram1997collision}, thus a finite collision rate in our simulations implies that the value of RDF should be equal to a fixed value but not an infinitesimal value. The results shown above agree with this analysis.}

{We shall call the region where the RDF decreases, as seen in Fig.~\ref{Fig.rdf_small_scale}, the depletion zone. Subsequent discussions will mainly focus on this region.}

\subsection{The Collisional Modulation Factor}

\begin{figure}[htb]
	\includegraphics[width=0.45\textwidth]{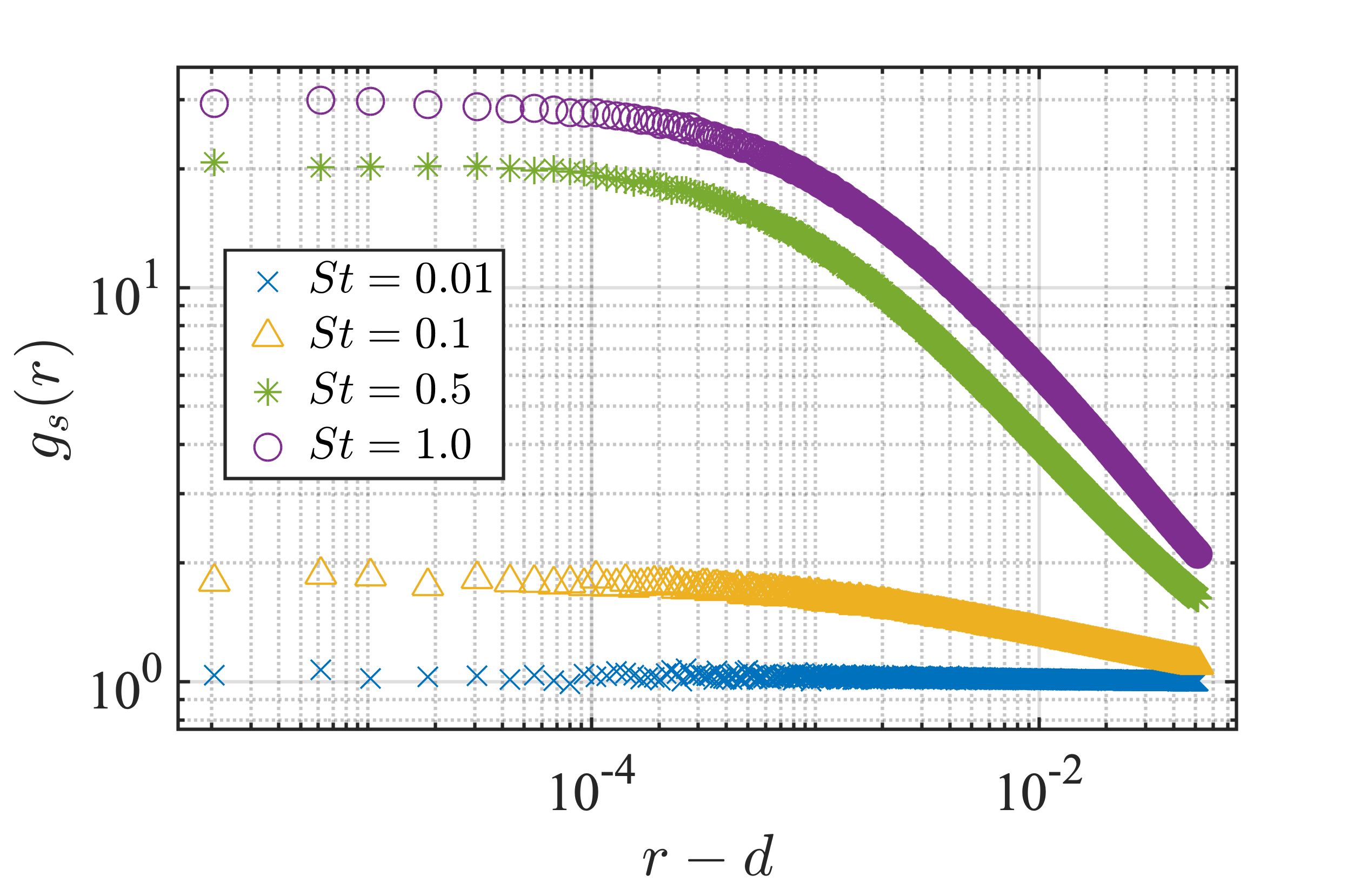}
	\caption{The RDF for non-colliding particles ($g_{n}$) for different Stokes numbers. The particle diameter is $d=9.49\times10^{-4}$ and $Re_{\lambda}=124$. In log-log axes, $g_n$ level off to plateau as the gap-distance $r-d$ approaches zero.}
	\label{Fig.gs_st}
\end{figure}

\begin{figure}[htb]
	\subfigure[]{
		\label{Fig.g0_st_part1}
		\includegraphics[width=0.45\textwidth]{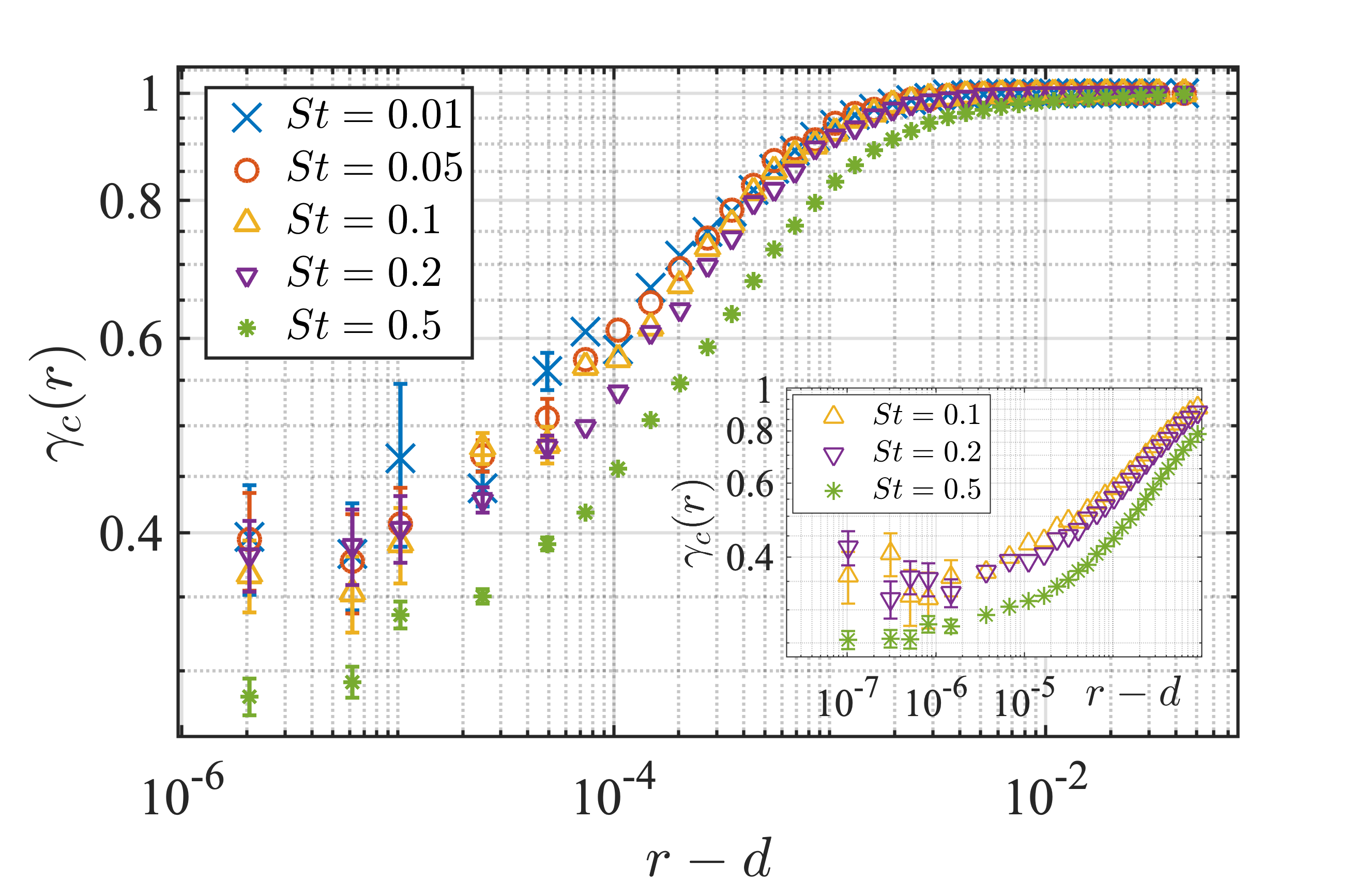}
	}
	\subfigure[]{
		\label{Fig.g0_st_part2}
		\includegraphics[width=0.45\textwidth]{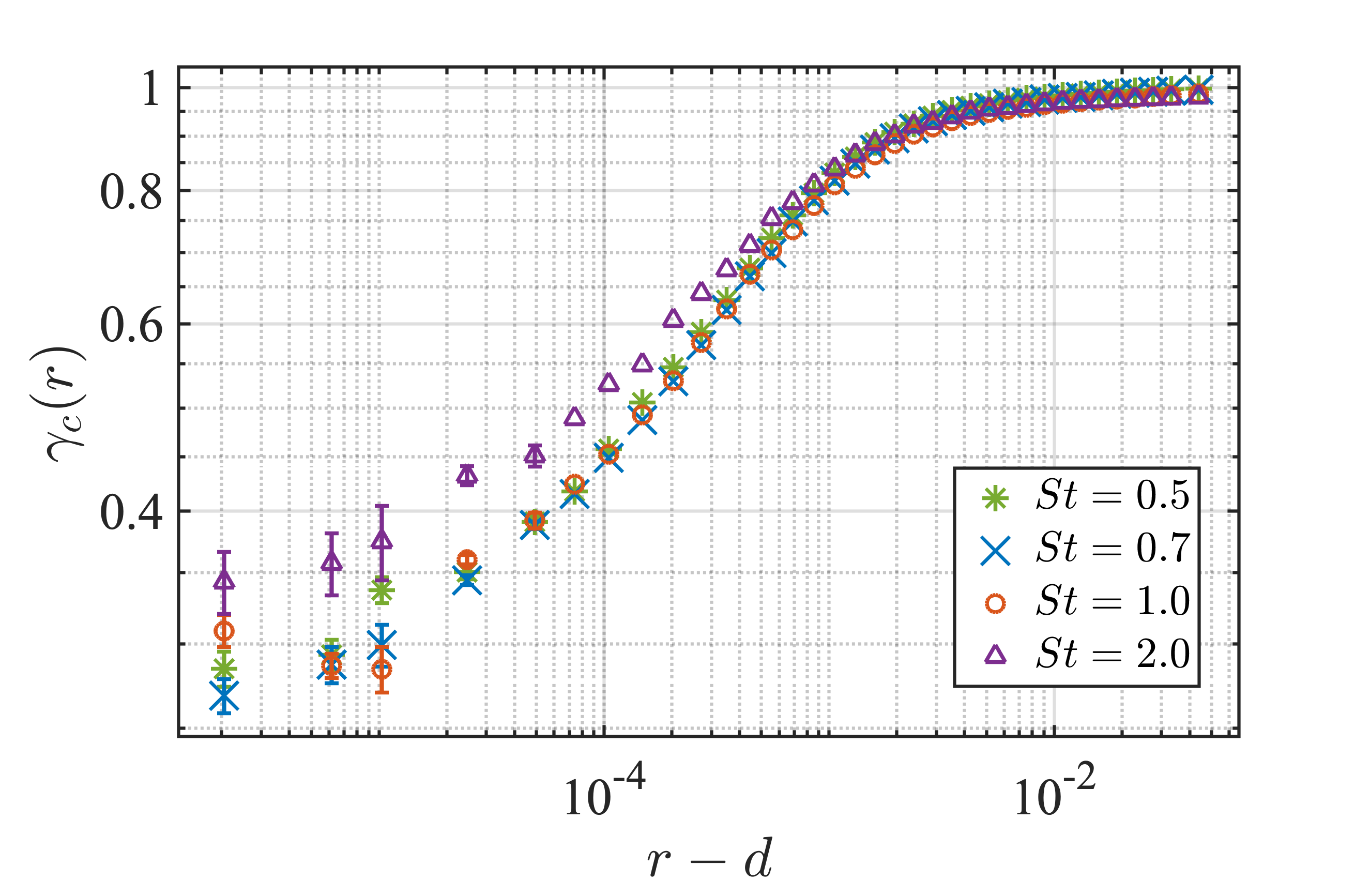}
	}
	\caption{The collisional modulation factor $\gamma_{c}$ versus the gap-distance $r-d$ for particles with different Stokes numbers. The particle diameter is $d=9.49\times10^{-4}$ and $Re_{\lambda}=124$. Within the range of uncertainty, which is calculated as the standard error, $\gamma_c$ is weakly dependent on $St$ for $St\ll1.0$ and it decreases as $St$ increases from around 0.1 to 0.5.}
	\label{Fig.g0_st}
\end{figure}

\ew{To further characterize this ``depletion zone" of RDF due to particle collisions, following the work of (Saw and Meng, 2022) \cite{saw2022intricate}, we assume that the RDF could be factorized such that $g(r)=\gamma_{c}(r)g_{n}(r)$, where $\gamma_{c}(r)$ represent the effect of collision-coagulation in the form of a ``modulation factor", while $g_{n}(r)$ is the RDF for non-colliding (ghost) particles under the same physical environment. It is well known that, for mono-dispersed particle, $g_{n}(r)$ is a power-law of $r$ \cite{saw2012spatial}\cite{chun2005clustering}. However, as shown in Figure \ref{Fig.gs_st}, when plotted against $r-d$, $g_{n}$ level off to plateau as $r$ decrease towards $d$ as a result of finiteness of $g_{n}(d)$. The collision induced modulation factor $\gamma_{c}$ which, by definition, equals $g(r)$ compensated (divided) by $g_n(r)$, is calculated as such in each case and shown in Figure \ref{Fig.g0_st}. As expected, at large $r$, $\gamma_{c}$ universally converge to unity, signifying that collisional effect are only appreciable at $r \sim d$. In the $r \sim d$ regime on the other hand, we see that as $St$ increases from a minute value (i.e., $0.01$), $\gamma_{c}$ gradually decreases, with smaller $r$ affected more strongly. The rate of this decrease (with respect to $St$) is at first very weak, consistent with the hypothesis in \cite{saw2022intricate} that $\gamma_c$ is independent of $St$ in the limit of small $St$. The observed rate of decrease becomes pronounced when $St$ increases from $0.1$ to $0.5$. Beyond $St=0.5$, $\gamma_{c}$ seem almost stagnant again until it reverse the trend and start to increases significantly when $St$ is greater than $1.0$.} 
\ew{This implies that when the Stokes number of particles is much smaller than 1.0, the influence of collision on the RDF is insensitive to the Stokes number. As the Stokes number increases, the influence of collision grows and peaks at $St=1.0$, which is a trend similar to the one of the power ($c_1$) of inertial clustering \cite{saw2012spatial}.}
\ew{In order to see the trend of $\gamma_c$ at small $r$ more clearly, using the case of $St=0.1$, 0.2, and 0.5 as examples, we plot in inset of Figure \ref{Fig.g0_st_part1}, $\gamma_c$ versus the gap distance $r-d$. Again, we see that $\gamma_c$ follows a semi-power-law for $r-d$ in the range of  $1-6\times10^{-4}$. At smaller $r-d$, the curves flatten and level off to a finite value. }

\begin{figure}[htb]
\includegraphics[width=0.45\textwidth]{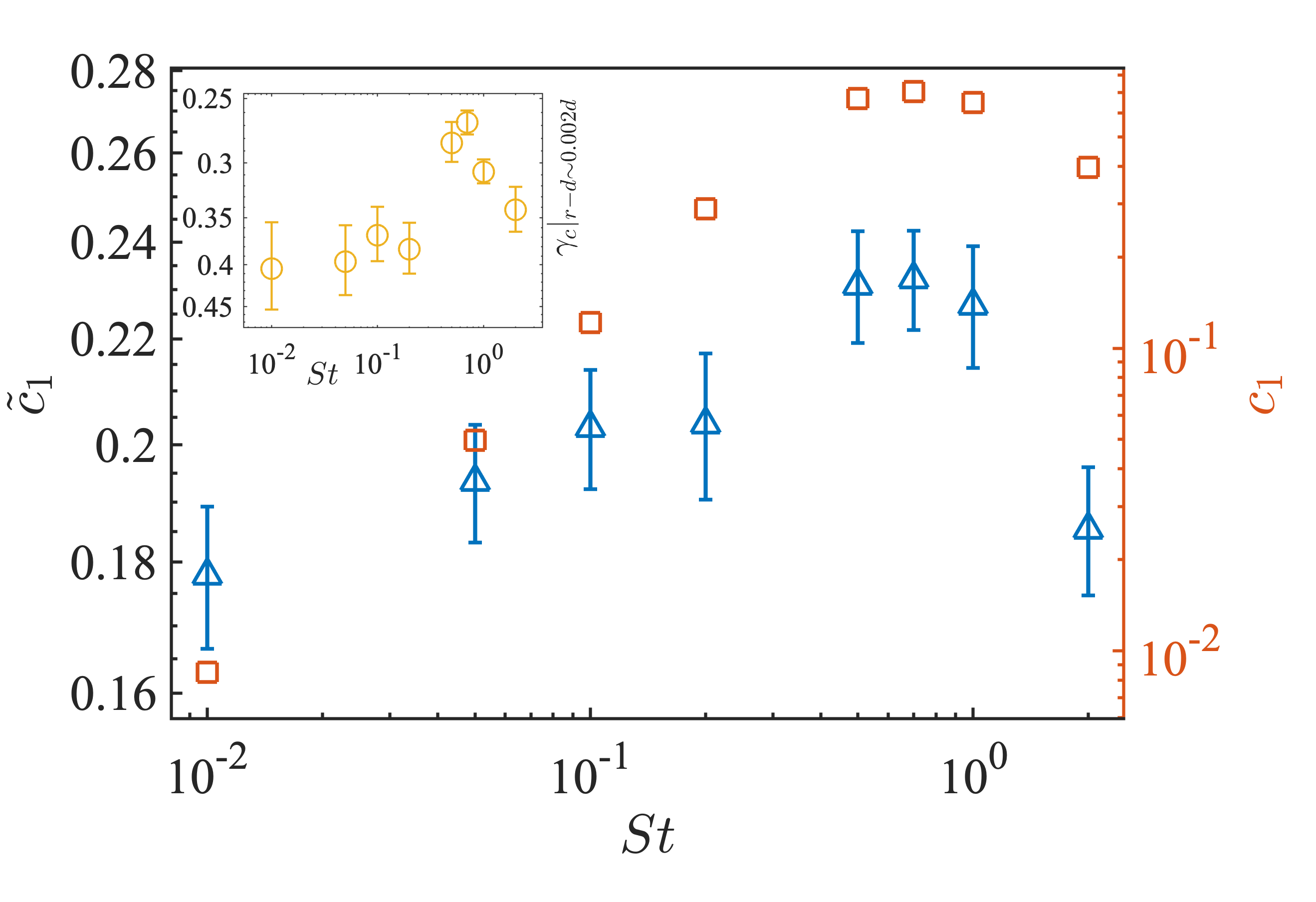}
\caption{The value of exponent of the power-law $\tilde{c}_1$ in $\gamma_c = \tilde{c}_{0}(r-d)^{\tilde{c}_1}$ in the range $1\times10^{-4}\lesssim r-d\lesssim7\times10^{-4}$, which is shown as \xh{blue $\triangle$}. Its error bars represent the statistical uncertainty in the nonlinear regression method. The value of $\gamma_c$ at $r\to d$ ($r-d\sim0.02d$) \ew{is shown as} \xh{red $\square$} and its error bars represent \ew{one} standard error. {\bf Inset)} The value of \ew{the} exponent of the power-law $c_1$, as defined in $g_{n}=c_{0}(r/eta)^{-c_{1}}$, obtained from DNS results in the range $0.1\lesssim r/\eta \lesssim 1$. The statistical uncertainty of $c_1$ is smaller than the size markers, therefore it is not shown in this figure. \xh{The vertical axes and the horizontal axis are logarithmic.} \xh{$\tilde{c}_1$ and $\gamma_{c}(r\to d)$ increase slowly for $St\ll 1$ and it grows significantly from $St=0.2$ to 0.5.} The general trend of $\tilde{c}_1$ and $-\gamma_c(r\to d)$ as a function of $St$ is similar to that of $c_1$.}
\label{Fig.gc_evaluate}
\end{figure}

\ew{
For the sake of comparison, let us recall from earlier works \cite{reade2000effect}\cite{chun2005clustering}\cite{saw2012spatial} that $g_{n}$ is a negative power law of $r$ in the regime of $r/\eta \lesssim 20$, i.e., $g_n(r)=c_{0}(r/\eta)^{-c_{1}}$, where $c_{1}$ scales as $St^2$ for $St\ll1$. We now attempt to derive an analogous relationship between $\gamma_{c}$ and the Stokes number. From Figure \ref{Fig.g0_st}, $\gamma_{c}$ seems to follow a quasi-power-law for gap distances $(r-d)$ in the range $10^{-4} \lesssim r-d \lesssim7\times 10^{-3}$. We assume that in this region $\gamma_{c} = \tilde{{c}}_{0}(r-d)^{\tilde{c}_1}$.
The relationship between $\tilde{c}_1$ and $St$ is shown in Figure \ref{Fig.gc_evaluate}. For $St\ll1$, considering the level of statistical uncertainty, there is a tentatively trend of increasing $\tilde{c}_1$ with Stokes number \xh{($\tilde{c}_1$ follows a quasi-linear relationship with $St$ for $St\ll 1$ in linear axes)}. This positive trend is clearer at larger $St$. $\tilde{c}_1$ reach a peak value at around $St=0.7$ and thereafter decreases slowly at larger $St$.} 

\ew{For comparison, we also show $c_1(St)$ and $\gamma_c(r\to d)$ in Figure \ref{Fig.gc_evaluate}, where $\gamma_c(r\to d)$ is the limiting value of $\gamma_c$ at particle contact ($r=d$).} \ew{ The latter is of interest since it is closely related to the collision rate (in practice, we use the value of $\gamma_c$ at $r-d\sim2\times10^{-6}$ as this limit). Comparing these three plots, we see that the general trend of both $\tilde{c}_1$ and $-\gamma_c(r\to d)$ as a function of Stokes number is similar to that of $c_1$.}

\ew{The above results indicate that when $St\ll1.0$, the drop of RDF caused by particle collision-coagulation is roughly independent of the Stokes number. While for large $St$ ($>0.2$), the RDF in depletion zone is related to it. Furthermore, the relationship between the decrease of RDF and $St$ is similar to that between the preferential concentration of particle and $St$. This implies that the Stokes number dependence of the RDF could not be complete decoupled from $\gamma_c$ and that this issue merit further investigations.}

\subsubsection{Comparative study of the full RDF}

\begin{figure}[htb]
\includegraphics[width=0.45\textwidth]{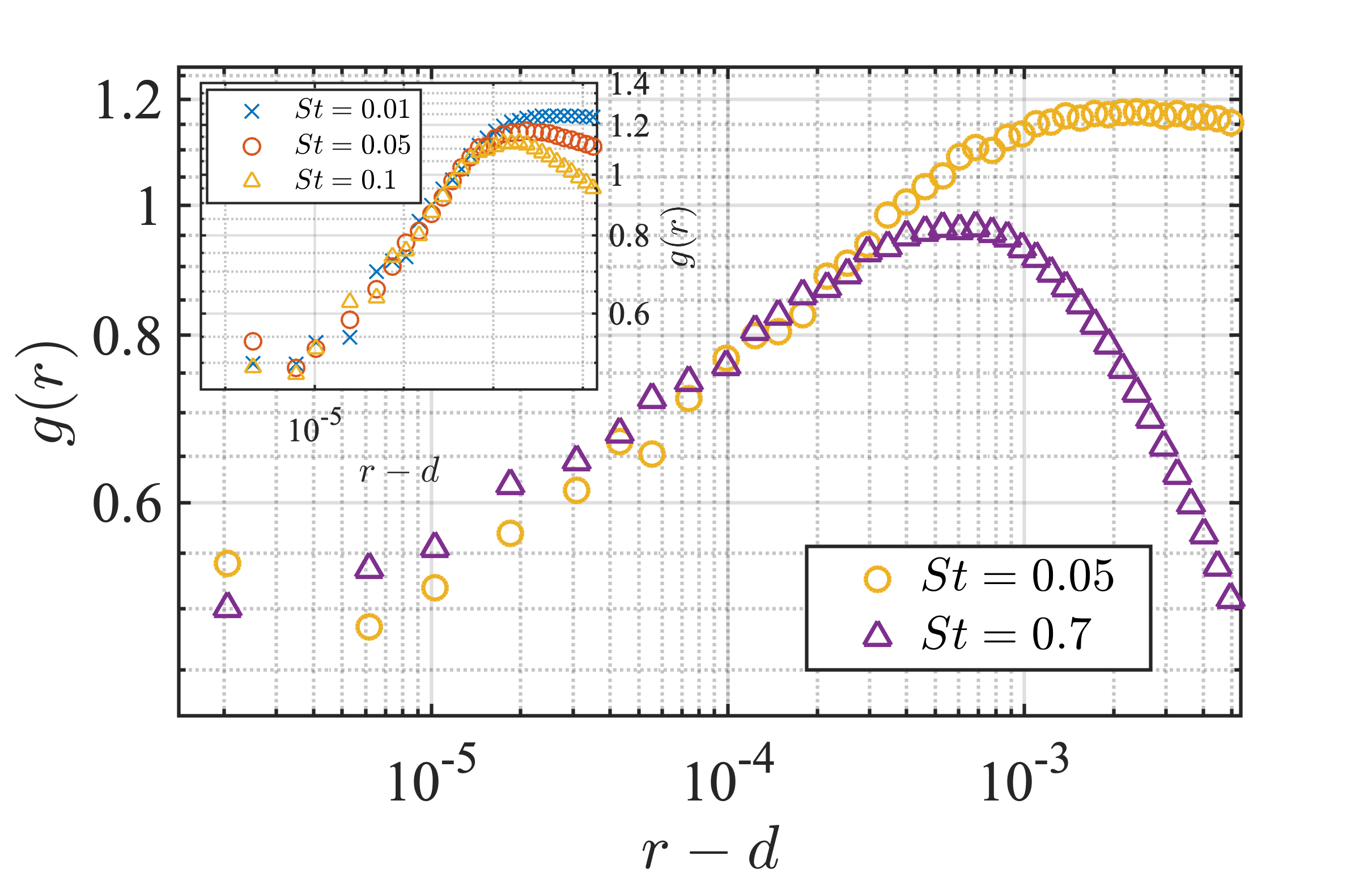}
\caption{\xh{The RDF for $St=0.7$ is translated vertically down to $6.5\%$ of its original height to compare the slope with $St=0.05$.} The translated RDFs for $St=0.01$, 0.05,  and 0.1 are in the inset. The slope is the same for $St\ll1$ but it is different for $St=0.7$ and 0.05. The position where the peak of RDF is related to $St$.}
\label{Fig.rdf_rescaled}
\end{figure}

\ew{In view of the significant Stokes number dependence of $\gamma_c$, we also conducted a complementary investigation on the Stokes number trend of the full RDFs (i.e., $g(r)$) as a comparative study.
The RDFs for various Stokes numbers are translated in vertically to overlap with the $g(r)$ for $St=0.05$ at $r-d\sim2\times10^{-4}$ to compare their shape. In order to show the influence on $St$ more clearly, the RDFs for $St=0.05$ and 0.7 are shown in Figure \ref{Fig.rdf_rescaled} and the RDFs for $St\ll1$ are shown in the inset. It can be seen that the slope of RDF is almost constant for $St\ll1$ but is changing when considering larger $St$. Furthermore, the position of the peak of the RDF has a significant $St$-dependence. This shows that both $\gamma_c(r)$ and $g(r)$ has nontrivial Stokes number dependence which should be further investigated in future works.}

\subsection{Reynolds number dependence}
\begin{figure}[htb]
\includegraphics[width=0.45\textwidth]{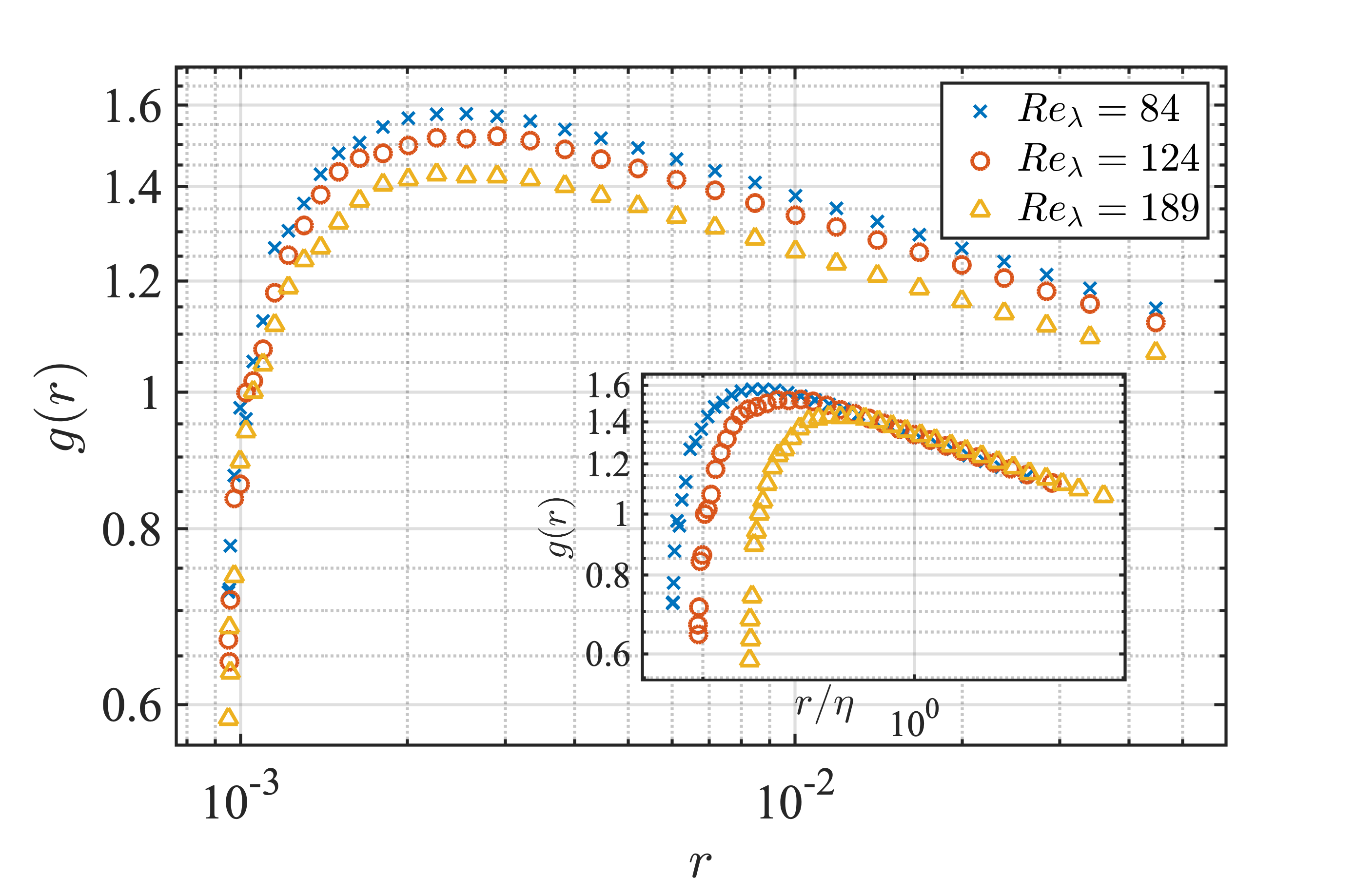}
\caption{The RDFs for particles in three cases with different $Re_{\lambda}$. \ew{The particle Stokes number is 0.1 and the diameter of particle is $d=9.49\times10^{-4}$. {\bf Inset)} the RDFs as the function of $r$ normalized by the Kolmogorov length $\eta$. The observed overlaps implies that Renoylds number effect is very weak.}}
\label{Fig.rdf_Re}
\end{figure}

Statistics of \ew{Run 3, Run 11, and Run 12, listed in Table \ref{tab2}}, are used to investigate the influence of {the Taylor-scaled} Reynolds number $Re_{\lambda}$ on RDF. The RDFs for {different $Re_{\lambda}$} are shown in \ew{Figure \ref{Fig.rdf_Re}. The Stokes number is 0.1 and the parameters for three simulations are shown in Table \ref{tab1}. We see that in the range of small $r$, the RDFs of all cases overlap but for larger $r$, the RDFs are separated. However, if $r$ is normalized by the Kolmogorov length scale $\eta$, as shown {in the inset of} Figure \ref{Fig.rdf_Re}, the RDFs now overlap at large $r$.} 
{These results suggest that in the range of $Re_{\lambda}=84-189$, the statistics of the depletion zone related to particle collision is not influenced by $Re_{\lambda}$, while the power regime related to the classical inertial clustering depends on $Re_{\lambda}$ but only through the change in Kolmogorov length-scale $\eta$.}

\begin{figure}[htb]
	\includegraphics[width=0.45\textwidth]{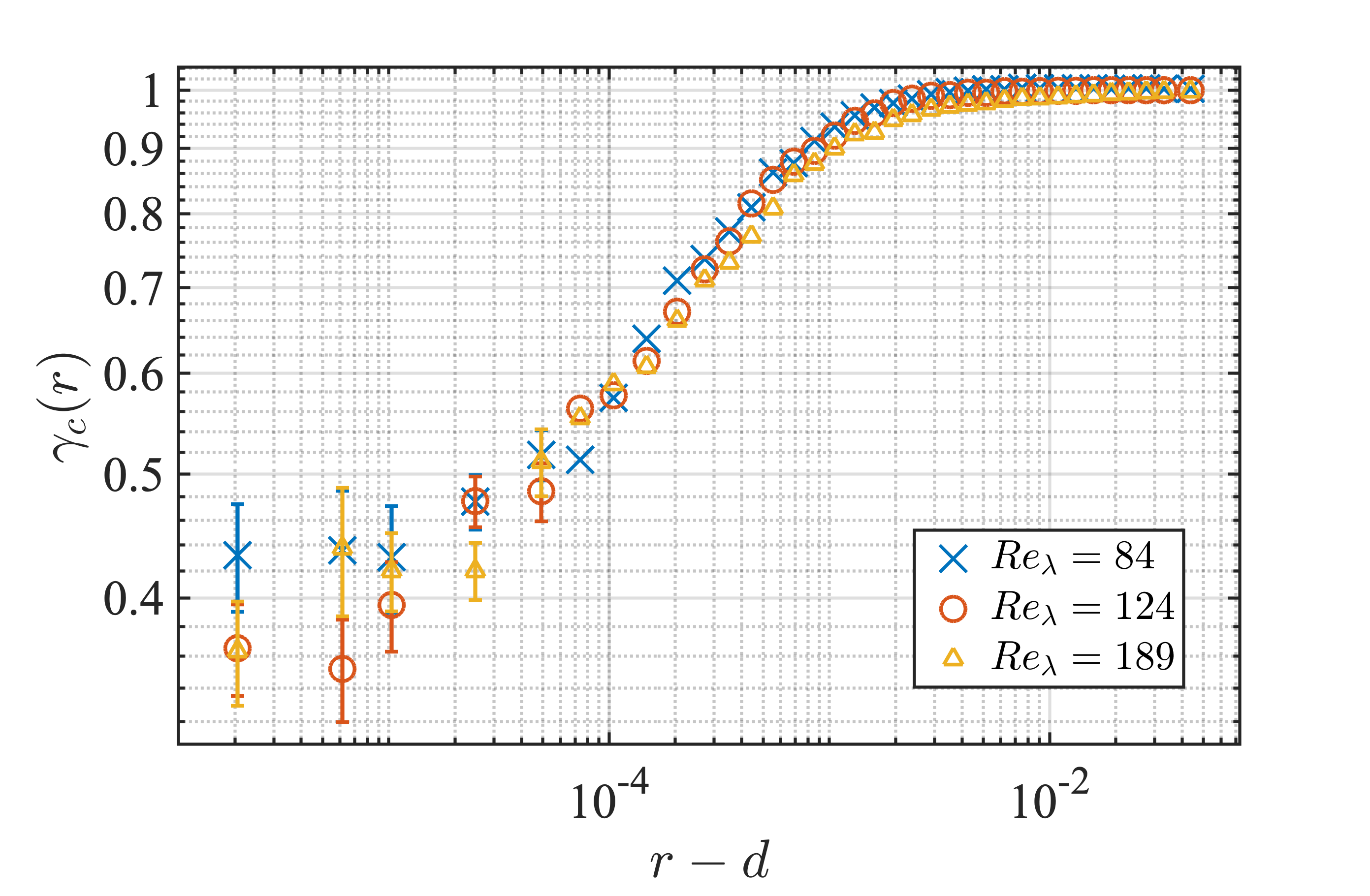}
	\caption{The modulation factor $\gamma_{c}$ versus $r-d$ for particles in three cases with different $Re_{\lambda}$. The Stokes number is 0.1 and the particle diameter $d=9.49\times10^{-4}$. $\gamma_c$ for three cases are overlapped within a range of uncertainty. Error bars represents standard deviation.}
	\label{Fig.g0_Re}
\end{figure}

\ew{The modulation factor $\gamma_{c}$ for these cases are shown in Figure \ref{Fig.g0_Re}. We find that they overlap within statistical uncertainty. {This again suggests that the main characteristics of the depletion zone is insensitive to flow Reynolds number from $Re_{\lambda}=84$ to 189.}} 

According to the Kolmogorov 1941 hypothesis \cite{kolmogorov1941local}, when the Reynolds number is large enough, the statistics of the small-scale of turbulent flow will not be influenced by of the large-scale. 
Collision occurs on a scale of particle diameter, which is here much smaller than the Kolmogorov length scale, therefore, the decrease of RDF is insensitive to $Re_{\lambda}$. This view is consistent with the above findings.

\subsection{Particle diameter dependence}

\begin{figure}[htb]
	\includegraphics[width=0.45\textwidth]{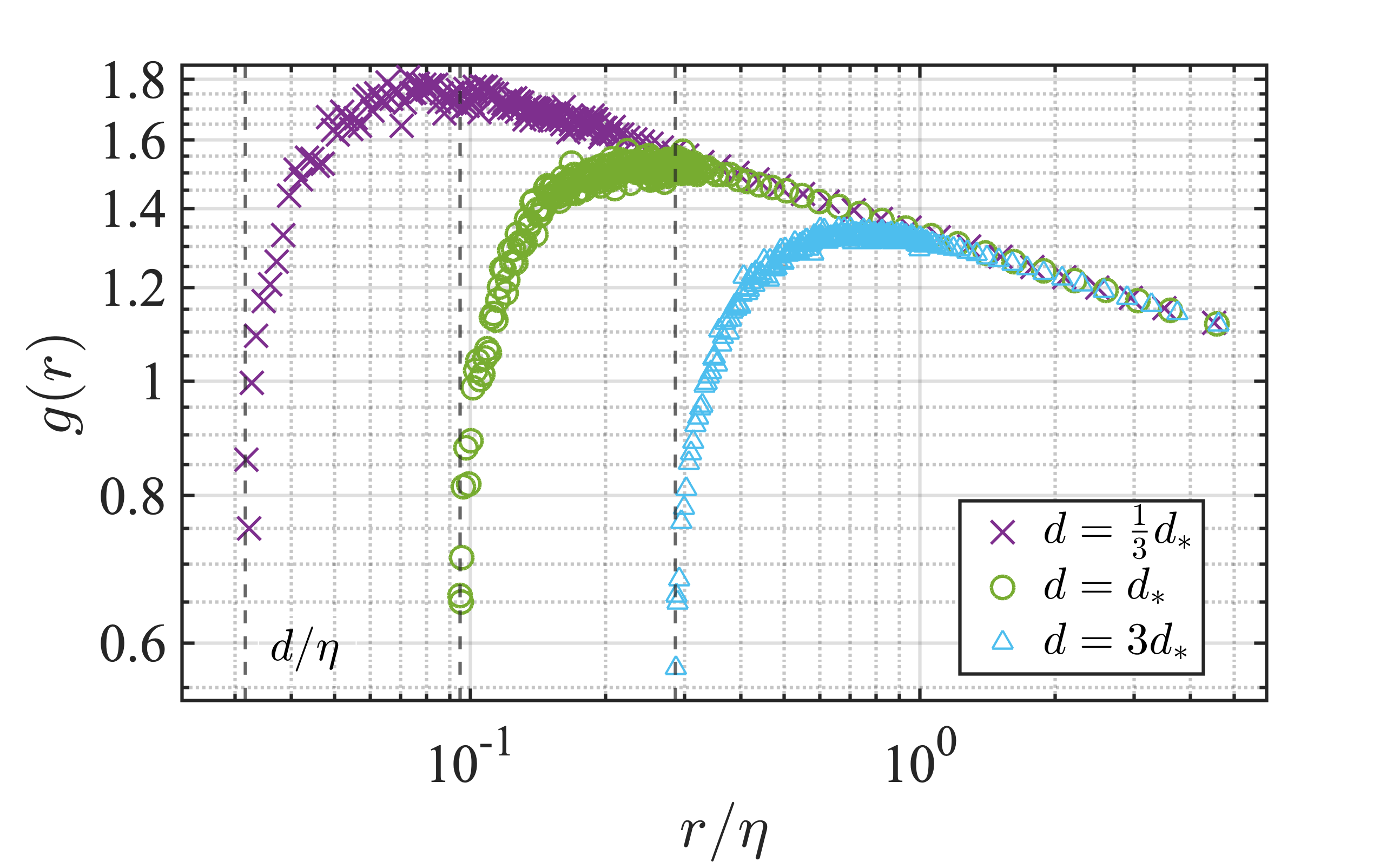}
	\caption{The RDFs for particles with different diameters, $d=\frac{1}{3}d_{*}$, $d=d_{*}$ and $d=3d_{*}$ respectively, in which $d_{*}=9.49\times10^{-4}$. The Stokes number of particles is 0.1 and the Reynolds number of the flow is $Re_{\lambda}=124$. \ew{The position where the RDF decreases closely follows to the particle diameter.}}
	\label{Fig.rdf_size}
\end{figure}

The RDFs for particles with different diameters are shown in Figure \ref{Fig.rdf_size}, the statistics used are from Run 3, Run 9, and Run 10. 
What is striking in Figure \ref{Fig.rdf_size} is that the position where RDF starts decreasing is consistent with particle diameter.

\begin{figure}[htb]
	\includegraphics[width=0.45\textwidth]{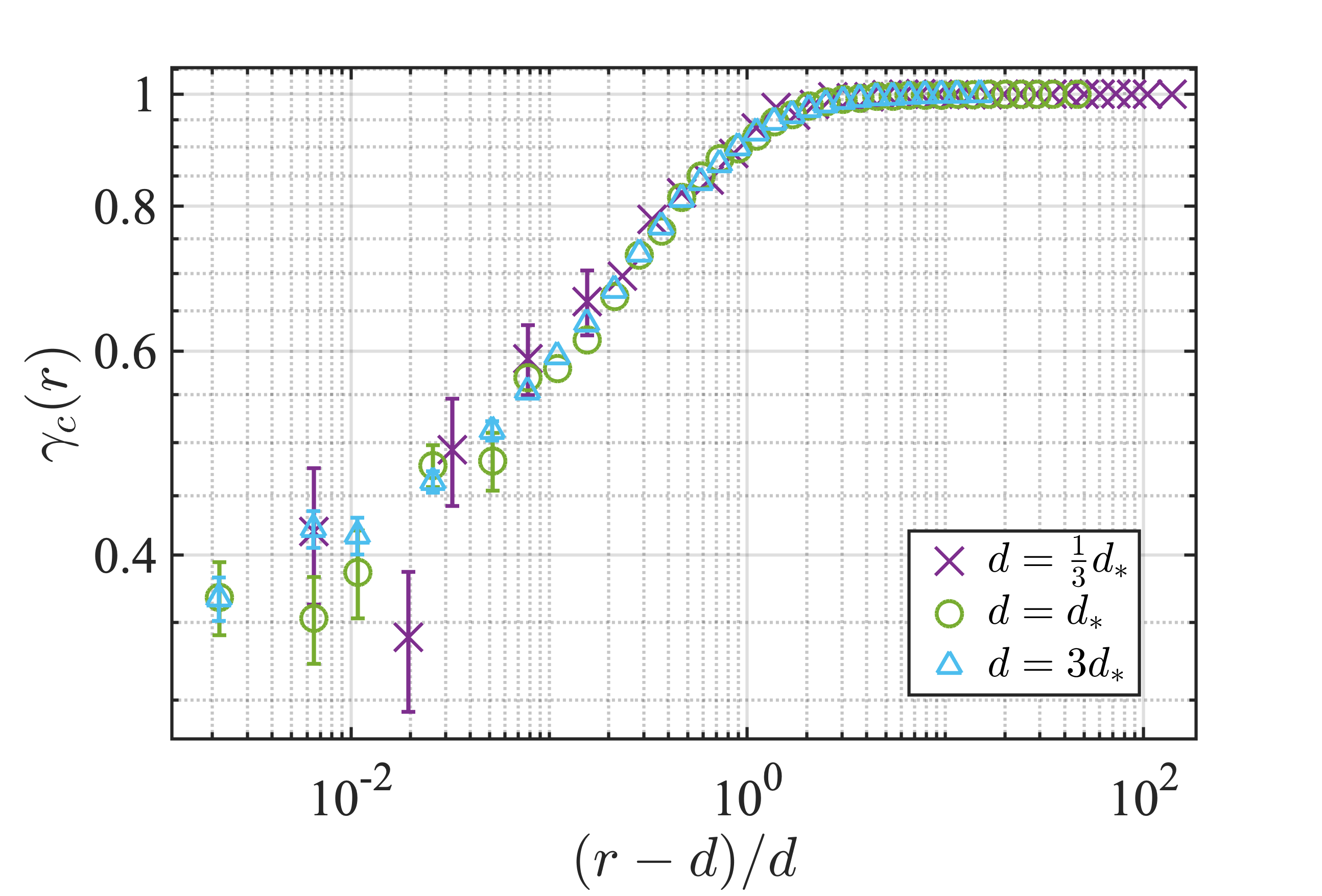}
	\caption{The collisional modulation factor $\gamma_{c}$ for particles with different diameters, $d=\frac{1}{3}d_{*}$, $d=d_{*}$ and $d=3d_{*}$, in which $d_{*}=9.49\times10^{-4}$. The Stokes number of particles is 0.1 and the Reynolds number of turbulent flow is $Re_{\lambda}=124$. $\gamma_c$ for all three cases coincide substantially.}
	\label{Fig.g0_size}
\end{figure}

The modulation factor $\gamma_{c}$ for particles with different diameter as a function of the rescaled gap distance $(r-d)/d$ is shown in Figure \ref{Fig.g0_size}. In this case, the modulation \xh{factors} $\gamma_{c}$ are \xh{coincident}.

\ew{The collision process of particles is strongly related to the particle size. Since we only consider the RDFs of mono-dispersed particles, we expected that the position where depletion zone begins is close to the particle diameter. The results shown here are in line with our expectations.}


\section{Conclusion}
\ew{This paper studies the change of radial distribution function (RDF) of particles subjected to the collision and coagulation (coalescence) interaction. We investigate the relationship between the RDF and particle Stokes number, particle diameter, and Reynolds number. We find that the RDF diminishes significantly at small particle separation distances $r$. When viewed as a function of $r-d$, we found evidence that the RDF do not decrease infinitely, but levels off to a fixed value in the limit of $r \to d$. 
To study the relationship between the degree of reduction of the RDF and the particle and turbulent parameters, we separate the RDF into two multiplicative parts i.e., $g=\gamma_{c}\,g_{n}$, where $g_{n}$ is the RDF for non-colliding particles under equivalent environment and $\gamma_{c}$ is a collisional modulation factor that reflects the effect of particle collision on particle preferential concentration. 
We see that \ew{$g_{n}(r-d)$} levels off to a plateau as the  argument $r-d$ approaches zero. \ew{On the other hand,} the collisional factor $\gamma_c$ universally converges to unity at large $r$ and levels off to a fixed value at $r\approx d$. \ew{We find that $\gamma_c$ is dependent on the Stokes number. Specifically, assuming a power-law model for $\gamma_c$ in the region $0.1d\lesssim r-d \lesssim 0.7d$ (i.e., $\gamma_c=\tilde{c}_{0}(r-d)^{\tilde{c}_{1}}$), we find that in the small Stokes number limit, the value of $\tilde{c}_{1}$ is only very weakly dependent on $St$, while the overall Stokes number trend of $\tilde{c}_1$ is qualitatively similar to the power law exponent $c_{1}$ in the RDF of non-colliding (ghost) particles (i.e., $g_{n}(r)
)$. The magnitude of $\gamma_c$ at the limit $r\to d$ varies with the Stokes number following a trend similar to that of $\tilde{c}_1(St)$.}}

{The preceding findings motivate a comparative investigation into the $St$ trend of the full RDF, which has the result that the slope of the RDF $g(r)$ in the depletion zone is the same for $St\ll1$ but it is different for large Stokes number (i.e., $St \geq 0.05$). Besides this, the location of the peak of RDF is found to be significantly $St$-dependent. These findings imply that the Stokes number dependence of the RDF could not be completely decoupled from $\gamma_c$ (nor from $g(r)$) except in the regime of $St\ll1$ where $St$ dependence is weak.}

\ew{We find that the shape of the RDF in the depletion zone ($r \sim d$) do not change with the variation of the flow Reynolds number within the range studied i.e., $Re_{\lambda}=84-189$ and the collisional modulation factor $\gamma_c$ from different $Re_{\lambda}$ overlap.}

On the effect of particle diameter $d$, we find that larger $d$ leads to the falling edge in $\gamma_c$ occurring at larger values of $r-d$ (and $r$) such that the results coincide when $\gamma_c$ is plotted against $(r-d)/d$.


\section*{Acknowledgements.} This work was supported by the National Natural Science Foundation of China (grant 11872382) and by the Thousand Young Talents Program of China.

\bibliography{rdf_references}

\end{document}